\documentclass[11pt,a4paper]{article}
\pdfoutput=1
\usepackage{exscale,times}
\usepackage{fancyhdr}
\usepackage[cp1250]{inputenc}
\usepackage[T1]{fontenc}
\usepackage{amsmath}
\usepackage{fancyhdr}
\usepackage{indentfirst}
\usepackage{times}
\usepackage{graphicx}
\usepackage{multicol}
\usepackage{here}
\usepackage{amsmath}
\usepackage{amssymb}
\usepackage{amsthm}
\usepackage{amsfonts}
\usepackage{stmaryrd}
\usepackage{mathrsfs}
\usepackage{color}
\usepackage{multicol}
\usepackage{mathrsfs}
\usepackage{wasysym}
\usepackage[yyyymmdd,hhmmss]{datetime}

\usepackage{sidecap}
\usepackage{textcomp}
\usepackage{enumitem}
\usepackage{upgreek}
\usepackage[normalem]{ulem}

\newcommand{\bup}{\mbox{\boldmath $\upsilon$}}

\newcommand{\bsigma}{\mbox{\boldmath $\sigma$}}

\newcommand{\bveps}{\mbox{\boldmath $\varepsilon$}}

\newcommand{\mytitle}{Fractional Euler-Bernoulli beams: theory, numerical study and experimental validation}

  \usepackage{hyperref}
 \hypersetup{
      bookmarks=true,         
      pdftoolbar=true,        
     pdftitle={\mytitle},    
      pdfauthor={Wojciech Sumelka, Tomasz Blaszczyk, Christian Liebold},     
      colorlinks=true,       
      linkcolor=blue,          
      citecolor=blue,        
  	bookmarksnumbered=true
 }

\begin{document}
\pagestyle{fancy}

\chead{Wojciech Sumelka, Tomasz Blaszczyk, Christian Liebold - Compiled on \today\ at \currenttime}
\lhead{}
\rhead{}
\renewcommand{\headrulewidth}{0.4pt}
%
%
\begin{center}
{\fontsize{14}{20}\bf{\mytitle}}
\end{center}

\begin{center}
\textbf{W. Sumelka$^*$, T. Blaszczyk$^{**}$, C. Liebold$^{***}$}\\
\bigskip
{$^*$Poznan University of Technology, Institute of Structural Engineering, }\\
Piotrowo 5 street, 60-969 Poznan Poland\\ 
wojciech.sumelka@put.poznan.pl\\
~\\
$^{**}$Czestochowa University of Technology, Institute of Mathematics\\
al. Armii Krajowej 21, 42-201 Czestochowa, Poland\\
tomasz.blaszczyk@im.pcz.pl\\
~\\
$^{***}$Berlin University of Technology, Institute of Mechanics\\
Einsteinufer 5, 10587 Berlin, Germany\\
christian.liebold@tu-berlin.de\\
\bigskip
\end{center}
%
%


\vspace{1cm}

\begin{center}
\textbf{ABSTRACT}\\[1mm]
\end{center}
{In this paper the classical \textsc{Euler-Bernoulli} beam (CEBB) theory is reformulated utilising fractional calculus. Such generalisation is called \textit{fractional Euler-Bernoulli beams} (FEBB) and results in non-local spatial description. The parameters of the model are identified based on AFM experiments concerning bending rigidities of micro-beams made of the polymer SU-8. In experiments both force as well as deflection data were recorded revealing significant size effect with respect to outer dimensions of the specimens. Special attention is also focused on the proper numerical solution of obtained fractional differential equation.} 

\section{Introduction}

A quantitative understanding of a size effect in engineering materials is of great importance in the design phase of micro- and nano-electromechanical systems (MEMS and NEMS), and as a technique of homogenisation regarding materials with real micro-structure. A size effect is reflected, for example, in a stiffer elastic response to external forces on small scales. This has been recognised in several experiments on metals and polymers, for example on copper (\textsc{Fleck} et al. 1994 \cite{Fleck1994}), silver (\textsc{Ma} and \textsc{Clarke} 1995 \cite{Ma1995}), zinc oxide (\textsc{Stan} et al. 2007 \cite{Stan2007}), lead (\textsc{Cuenot} et al. 2004 \cite{Cuenot2004}), carbon nanotubes (\textsc{Salvetat} et al. 1999 \cite{Salvetat1999}), epoxy (\textsc{Chong} 2002 \cite{Chong2002}) and polypropylene (\textsc{McFarland} and \textsc{Colton} 2005 \cite{McFarland2005}). Actually, a physical reasoning for the origin of a size-dependent material behaviour, such as non-negligible complex interactions of molecular chains in polymers (\textsc{Nikolov} et al. 2007 \cite{Nikolov2007}), the rearrangement of atoms or molecules near the surface, the influence of the grain-size in polycystals (\textsc{Smyshlyaev} and \textsc{Fleck} 1996 \cite{Smy1996}) and long-ranging influences from dislocations, voids or some inner micro- or nano-structure, is manifold and poorly understood especially in combined effects. For the reason that conventional continuum theory is unable to predict size effect, different non-conventional continua are proposed in literature, like \textit{non-local theories} (\textsc{Peddieson} et al. 2003 \cite{Peddieson2003}, \textsc{Eringen} 2010 \cite{Eringen2010}), \textit{strain-gradient theories} (\textsc{Toupin} 1962 \cite{Toupin1962}, \textsc{Mindlin} and \textsc{Eshel} 1968 \cite{Mindlin1968}), \textit{micropolar theories} (\textsc{Eringen} 1966 \cite{Eringen1966}, \textsc{Nowacki} 1972 \cite{Nowacki1972}) or \textit{theories of material surfaces} (\textsc{Gurtin \& Murdoch} 1975 \cite{Gurtin1975}).

In addition to these, the \textit{fractional calculus} is assumed to be a promising candidate for modelling scale-dependent material behaviour too, by using a concept of intrinsic fractal structures to describe a potential disordered material's micro-structure (cf. \textsc{Carpinteri} 1994 \cite{Carpinteri1994}). {For detailed descriptions see} \textsc{Klimek} (2001) \cite{Klimek2001}, \textsc{Vazaquez} (2004) \cite{Vazaquez2004}, \textsc{Lazopoulos} (2006) \cite{Lazopoulos2006}, \textsc{di Paola} et al. (2009) \cite{Paola2009}, \textsc{Atanackovic \& Stankovic} (2009) \cite{Atanackovic2009}, \textsc{Carpinteri} et el. (2011) \cite{Carpinteri2011} or \textsc{Drapaca \& Sivaloganathan} (2012) \cite{Drapaca2012}. The formulation of fractional elasticity stated here defines spatial derivatives of arbitrary order including the positive features -- in contrast to the specified non-conventional continua~--~(1) to introduce a smaller number of additional material parameters, (2) to fit into the general framework of classical continuum mechanics and (3) to possess clear physical interpretation (e.g. regarding to the intrinsic length scale).

In Sect.~\ref{subsec:2.1}, we start with a description of a generalised continuum mechanical framework including the specific formalism of fractional calculus proposed in \textsc{Sumelka} 2014 \cite{Sumelka2013-TH}. More precisely, we define the gradient of the motion of a body, i.e. the deformation gradient, with the help of an differentiation operator a  of real order. Having this, strain and stress measures, as well as a formulation of the conservation of momentum will be derived according to conventional kinematics of small deformations. In Sect.~\ref{subsec:2.2} the classical \textsc{Euler-Bernoulli} assumptions will be utilised to define a displacement field valid for bending of slender beams. By applying equilibrium considerations to an infinitesimal beam element, the \textit{fractional} \textsc{Euler-Bernoulli} \textit{differential equation} will be presented as the basis for the numerical study. Section~\ref{subsec:3.1} clarifies the boundary conditions for the model in accordance with the experimental realisation. We will investigate the bending behaviour of micro-beams with rectangular cross-sections that are clamped on one side and loaded at the other by a single concentrated force. The computational algorithm, explained in more detail in Sect.~\ref{subsec:3.2}, will use a one-dimensional mesh of equidistantly distributed calculation points along the beam's axis. In between these nodes, the \textit{fractional} \textsc{Euler-Bernoulli} \textit{differential equation} will be solved including additional conditions for the transition of moments and forces between the sub-elements of the model. However, the operator for the differentiation of real order requires an additional step of extended integration over the neighbourhood of the relevant point to collect and weigh the non-local information of the body. The step of integration will be implemented using the fractional trapezoidal scheme \textsc{Odibat} (2006) \cite{Odibat2006}, \textsc{Blaszczyk} {et al.} (2013) \cite{Blaszczyk2013} and \textsc{Sumelka \& Blaszczyk} (2014) \cite{ Sumelka2014-AoM}.
All these preparations allow us to build a stiffness matrix of the system, to be solved by {the LU decomposition method.} 
The benchmark numerical results will be presented in Sect.~\ref{subsec:3.3} for beams of different (thicknesses/lengths) and different fractional parameters will be applied to optimise the results to fit to the experiments.

Finally, in Sect.~\ref{sec:4} with the help of an Atomic Force Microscope (AFM) force as well as deflection data of micro-beams made of the polymer material SU-8 the fractional beam model will be identified. The experimental approach and the measured data will be explained to emphasise that the bending stiffness of the SU-8 structures revealed a size effect on the thickness and the length of the samples.

\section{Fractional Euler-Bernoulli beams}\label{sec:2}
\subsection{Small strain fractional elasticity}\label{subsec:2.1}
There are many ways to generalise continuum mechanics in terms of fractional calculus, as mentioned in \textsc{Klimek} (2001) \cite{Klimek2001}, \textsc{Vazaquez} (2004) \cite{Vazaquez2004}, \textsc{Lazopoulos} (2006) \cite{Lazopoulos2006}, \textsc{di Paola} et al. (2009) \cite{Paola2009}, \textsc{Atanackovic \& Stankovic} (2009) \cite{Atanackovic2009}, \textsc{Carpinteri} et el. (2011) \cite{Carpinteri2011} or \textsc{Drapaca \& Sivaloganathan} (2012) \cite{Drapaca2012}. Herein we follow the concept presented in \textsc{Sumelka} (2014) \cite{Sumelka2013-TH}. Because the FEBB theory results from putting specific restrictions on general small strain fractional elasticity, we start with a short introduction to the latter one -- for details cf. \textsc{Sumelka} (2014) \cite{Sumelka2013-TH}, \textsc{Sumelka \& Blaszczyk} (2014) \cite{Sumelka2014-AoM}.

The concept of fractional continua discussed in  \textsc{Sumelka} (2014) \cite{Sumelka2013-TH} is based on fractional deformation gradients concept, namely 
\begin{equation}\label{eq:fracdefGradX}
\underset{X}{\tilde{\mathbf{F}}}(\mathbf{X},t)=\ell_X^{\alpha-1}\underset{X}{D}^\alpha\phi(\mathbf{X},t),\quad \mathrm{or}\quad {\underset{X}{\tilde{F}}} ~_{aA}=\ell_{A}^{\alpha-1}\underset{X_A}{D}^\alpha\phi_a\mathbf{e}_a\varotimes\mathbf{E}_A,
\end{equation}
and
\begin{equation}\label{eq:fracdefGradx}
\underset{x}{\tilde{\mathbf{F}}}(\mathbf{x},t)=\ell_x^{\alpha-1}\underset{x}{D}^\alpha\varphi(\mathbf{x},t),\quad \mathrm{or}\quad {\underset{x}{\tilde{F}}} ~_{Aa}=\ell_a^{\alpha-1}\underset{x_a}{D}^\alpha\varphi_A\mathbf{E}_A\varotimes\mathbf{e}_a,
\end{equation}
where $\underset{X}{\tilde{\mathbf{F}}}$ and $\underset{x}{\tilde{\mathbf{F}}}$ are fractional deformation gradients in material and spatial descriptions, respectively, $\ell_X$ and $\ell_x$ are corresponding length scales, $\phi$ defines the regular motion of the material body while $\varphi$ its inverse.  The fractional differential operator ${D}^\alpha$ is defined as the Riesz-Caputo (RC) fractional derivative
\begin{equation}\label{eq:RC}
~_{~~a}^{\mathrm{RC}}D^{\alpha}_b f(t)=\frac{1}{2}\frac{\Gamma(2-\alpha)}{\Gamma(2)}\left(~_{a}^{\mathrm{C}}D^{\alpha}_t f(t)+(-1)^n~_{t}^{\mathrm{C}}D^{\alpha}_b  f(t)\right),	
\end{equation}

where $\alpha\mkern-3mu>\mkern-3mu0$ denotes the real order of the derivative, $D$ denotes 'derivative' (RC stands for \textsc{Riesz-Caputo}), $a,t,b$ are so called terminals, $\Gamma$ is the \textsc{Euler} gamma function, $_{a}^{\mathrm{C}}D^{\alpha}_t f(t)$, $~_{t}^{\mathrm{C}}D^{\alpha}_b f(t)$ are left and right \textsc{Caputo}'s fractional derivatives, respectively \textsc{Podlubny} (1999) \cite{Podlubny1999-AP}, \textsc{Kilbas et al.} (2006) \cite{Kilbas06}, \textsc{Leszczynski} (2011) \cite{Leszczynski2011}. From the definition of the RC derivative by Eq.~(\ref{eq:RC}) it is clear, that fractional deformation gradients given by Eqs~(\ref{eq:fracdefGradX}) and (\ref{eq:fracdefGradx}) are non-local (each component of $\underset{X}{\tilde{\mathbf{F}}}$ and $\underset{x}{\tilde{\mathbf{F}}}$ governs the information from the surrounding, described by terminals $a$ and $b$). In consequence, other related measures of deformation (e.g. strain tensor) which are obtained through the relations analogous as in classical case, are also non-local. To conclude, fractional kinematics is controlled by new material parameters: length scale, order of derivation and type of fractional derivative \textsc{Oliveira \& Machado} (2014) \cite{Oliveira-2014} -- they should be identified for specific material.

\begin{figure}[H]
\centering
\includegraphics[width=10cm]{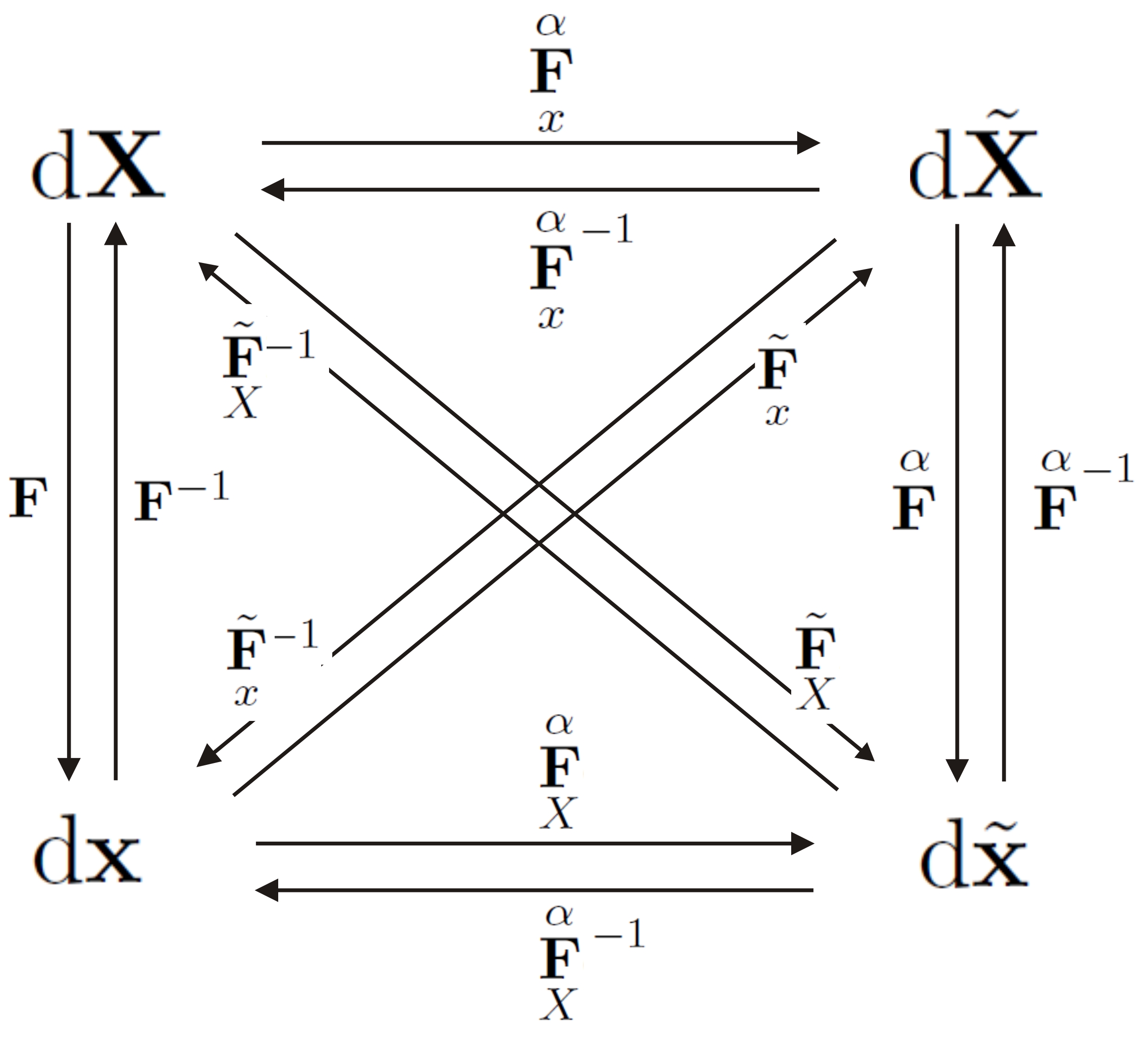}
\caption{The relationship between material and spatial line elements with their fractional counterparts $\overset{\alpha}{\mathbf{F}}=\underset{X}{\tilde{\mathbf{F}}}\mathbf{F}^{-1}\underset{x}{\tilde{\mathbf{F}}} ~^{-1}$, $\underset{x}{\overset{\alpha}{\mathbf{F}}}=\underset{x}{\tilde{\mathbf{F}}}\mathbf{F}$ and $\underset{X}{\overset{\alpha}{\mathbf{F}}}=\underset{X}{\tilde{\mathbf{F}}}\mathbf{F}^{-1}$.}
\label{fig:diagram}
\end{figure}

Thus, the finite fractional strains are obtained from the difference in scalar products in actual and reference configurations (cf. Fig.~\ref{fig:diagram})  

		\begin{equation}\label{eq:Efrac}
			\mathbf{E}=\frac{1}{2}(\overset{\Diamond}{\mathbf{F}} ~^T\overset{\Diamond}{\mathbf{F}}-\mathbf{I}),\quad \mathrm{or}\quad {E}_{AB}=\frac{1}{2}(\overset{\Diamond}{F} ~^T_{Aa}\overset{\Diamond}{F} ~_{aB}-{I}_{AB})\mathbf{E}_A\varotimes\mathbf{E}_B,
\end{equation}

		\begin{equation}\label{eq:efrac}			
		\mathbf{e}=\frac{1}{2}(\mathbf{i}-\overset{\Diamond}{\mathbf{F}} ~^{-T}\overset{\Diamond}{\mathbf{F}} ~^{-1}),\quad \mathrm{or}\quad {e}_{ab}=\frac{1}{2}({i}_{ab}-\overset{\Diamond}{F} ~_{aA}^{-T}\overset{\Diamond}{F} ~^{-1}_{Ab})\mathbf{e}_a\varotimes\mathbf{e}_b,
		\end{equation}

where $\mathbf{E}$ is the classical \textsc{Green-Lagrange} strain tensor or its fractional counterpart (symmetric), $\mathbf{e}$ is the classical \textsc{Euler-Almansi} strain tensor or its fractional counterpart (symmetric), and $\overset{\Diamond}{\mathbf{F}}$ can be replaced with $\mathbf{F}$ or $\underset{X}{\tilde{\mathbf{F}}}$ or $\underset{x}{\tilde{\mathbf{F}}}$ or $\overset{\alpha}{\mathbf{F}}$ (cf. Fig.~\ref{fig:diagram}). 

An infinitesimal fractional strain is obtained as in the classical set-up by considering the relationship between fractional strains and fractional displacement gradients tensors and omitting higher order terms in obtained relations. Hence, infinitesimal fractional \textsc{Cauchy} strain tensor is defined as (assuming that $\ell=\ell_X=\ell_x$)
\begin{equation}\label{eq:smallStr1D-2}
\overset{\Diamond}{\bveps}=\frac{1}{2}\left[\mathrm{Grad}\underset{X}{\tilde{\mathbf{U}}}
+\mathrm{Grad}\underset{X}{\tilde{\mathbf{U}}}^T\right]=\frac{1}{2}\left[\mathrm{grad}\underset{x}{\tilde{\mathbf{u}}}
+\mathrm{grad}\underset{x}{\tilde{\mathbf{u}}}^T\right],
\end{equation}
where
\begin{equation}
\mathrm{Grad}\underset{X}{\tilde{\mathbf{U}}}=\underset{X}{\tilde{\mathbf{F}}}-\mathbf{I} ,\quad \mathrm{or}\quad \ell^{\alpha-1}\underset{X_A}{D}^\alpha U_a=(\underset{X}{\tilde{{F}}} ~_{aA} - I_{aA})\mathbf{e}_a\varotimes\mathbf{E}_A,
\end{equation}
and
\begin{equation}
\mathrm{grad}\underset{x}{\tilde{\mathbf{u}}}=\mathbf{i}-\underset{x}{\tilde{\mathbf{F}}},\quad \mathrm{or}\quad \ell^{\alpha-1}\underset{x_a}{D}^\alpha u_A=(i_{Aa}-\underset{x}{\tilde{{F}}} ~_{Aa})\mathbf{E}_A\varotimes\mathbf{e}_a.
\end{equation}
In the above equations $\mathbf{U}$ and $\mathbf{u}$ denote material and spatial displacement.

It is clear that for the fractional kinematics appropriate fractional kinetics should be defined \textsc{Sumelka} et al. (2014) \cite{Sumelka2014-AAM}. It can be shown that fractional stress field $\tilde{\bsigma}$ satisfies analogous relations as in the classical formulation. Assuming that the conservation of mass holds, hence
\begin{equation}
{\rho_0}\mathrm{d}{V}=\tilde{\rho_0}\mathrm{d}\tilde{V}={\rho}\mathrm{d}{v}=\tilde{\rho}\mathrm{d}\tilde{v},
\end{equation}
or shortly
\begin{equation}
 \overset{\Diamond}{\rho_0}=\overset{\Diamond}{J}\overset{\Diamond}{\rho},
\end{equation}
where $\rho_0$ ($\tilde{\rho_0}$) is the reference mass density (fractional counterpart), $\rho$ ($\tilde{\rho}$) is spatial mass density (fractional counterpart), and $\overset{\Diamond}{J}=\mathrm{det}\overset{\Diamond}{\mathbf{F}}$ is a \textsc{Jacob}ian, we have
\begin{equation}
\int_{\overset{\Diamond}{v}}\overset{\Diamond}{\rho}\overset{\Diamond}{\dot{\bup}}\mathrm{d}\overset{\Diamond}{v}=\int_{\overset{\Diamond}{v}}\mathrm{div}\overset{\Diamond}{\bsigma} ~^{T}\mathrm{d}\overset{\Diamond}{{v}}+\int_{\overset{\Diamond}{v}}{\overset{\Diamond}{\rho}\mathbf{f}}\mathrm{d}\overset{\Diamond}{{v}},
\end{equation}
where $\bup$ is a velocity, and $\mathbf{f}$ is a body force per unit mass, so the local form is
\begin{equation}\label{eq:local}
\mathrm{div}\overset{\Diamond}{\bsigma} ~^{T}+{\overset{\Diamond}{\rho}\mathbf{f}}=\overset{\Diamond}{\rho}\overset{\Diamond}{\dot{\bup}},
\end{equation}
or in the absence of inertia forces
\begin{equation}
\mathrm{div}\overset{\Diamond}{\bsigma} ~^{T}+{\overset{\Diamond}{\rho}\mathbf{f}}=\mathbf{0}.
\end{equation}
Furthermore, the symmetry of $\overset{\Diamond}{\bsigma}$ (so $\bsigma$ or $\tilde{\bsigma}$) can be checked in a classical manner utilising the balance of moment of momentum. The result given by Eq.~(\ref{eq:local}) gives justification and is analogical for the one postulated in \textsc{Atanackovic \& Stankovic} (2009) \cite{Atanackovic2009} for similar small strain elasticity generalisation.

Finally, the small strain fractional elasticity is governed by
(for purpose the denotation $\tilde{(\cdot)}$ is omitted)
\begin{equation}\label{eq:govEqn}
\begin{cases}
\sigma_{ij,j}+b_i=0,\\
\overset{\Diamond}{\varepsilon}_{ij}=\frac{1}{2}\ell^{\alpha-1}\left(\underset{X_j}{D}^\alpha u_i+\underset{X_i}{D}^\alpha u_j\right)=\frac{1}{2}\ell^{\alpha-1}\left(u_{i,\breve{j}}+u_{j,\breve{i}} \right) ,\\
\sigma_{ij}=\mathcal{L}^e_{ijkl}\overset{\Diamond}{\varepsilon}_{kl},\\
U_i=\check{U}_i,\quad\quad \mathbf{X}\in \Omega_U,\\
\sigma_{ij}n_j=\check{t}_i,\quad\quad \mathbf{X}\in \Omega_\sigma,\\
\Omega_U\cap\Omega_\sigma=\emptyset\quad\mathrm{and}\quad\Omega_U\cup\Omega_\sigma=\Omega.
\end{cases} 
\end{equation}
In the above equation we have denoted: $\mathbf{b}$ is the body force, $\mathcal{L}^e$ is the stiffness tensor, $\Omega_U$ and $\Omega_\sigma$ are parts of boundary $\Omega$ where the displacements and the tractions are applied, respectively. It is clear that the classical (local) solution appears as a special case of the proposed generalisation when $\alpha=1$.

\subsection{Governing equations}\label{subsec:2.2}

Governing equations for the FEBB are obtained under the analogous assumptions as in a classical case (for classical case cf. {{\textsc{Bauchau \& Craig}} (2009)} \cite{Bauchau-2009}). Without loss of generality we consider the effect of extension and bending (other concepts like \textsc{Saint-Venant} assumptions associated with torsion can be incorporated by analogy).

The classical \textit{ad hoc} assumptions for kinematics hold, namely:
\begin{description}
\item[Assumption 1] The cross-section is infinitely rigid in its own plane,
\item[Assumption 2] The cross-section of a beam remains plane after deformation,
\item[Assumption 3] The cross-section remains normal to the deformed axis of the beam.
\end{description}
Within the above assumptions we can express the 3D displacements in terms of the 1D beam displacements (direction $1$ coincides with a beam axis):
\begin{eqnarray}\label{eq:dispAss}
\nonumber
u_1(x_1,x_2,x_3)&=&\bar{u}_1(x_1)+x_3\Phi_2(x_1)-x_2\Phi_3(x_1),\\
u_2(x_1,x_2,x_3)&=&\bar{u}_2(x_1),\\
\nonumber
u_3(x_1,x_2,x_3)&=&\bar{u}_3(x_1).
\end{eqnarray}
In Eqs~(\ref{eq:dispAss}) we have assumed the following sign convention: the rigid body translations of the cross-section $\bar{u}_1(x_1)$, $\bar{u}_2(x_1)$, and $\bar{u}_3(x_1)$ are positive in the positive direction of the coordinate axis, respectively; the rigid body rotations of the cross-section $\Phi_2(x_1)=-\frac{\mathrm{d}\bar{u}_3(x_1)}{\mathrm{d}x_1}$ and $\Phi_3(x_1)=\frac{\mathrm{d}\bar{u}_2(x_1)}{\mathrm{d}x_1}$ are positive keeping the right-hand rule.

Because finally the experimental set-up considers {bending in (1,3)-plane} we reduce the set of Eqs~(\ref{eq:dispAss}) to
\begin{eqnarray}\label{eq:dispAss2}
\nonumber
u_1(x_1,x_2,x_3)&=&-x_3\frac{\mathrm{d}\bar{u}_3(x_1)}{\mathrm{d}x_1},\\
u_2(x_1,x_2,x_3)&=&0,\\
\nonumber
u_3(x_1,x_2,x_3)&=&\bar{u}_3(x_1).
\end{eqnarray}
Applying the displacement relations Eqs~(\ref{eq:dispAss2}) to the definition of fractional \textsc{Cauchy} strain Eq.~(\ref{eq:smallStr1D-2}) we observe that the only non-zero element is
\begin{equation}
\overset{\Diamond}{\varepsilon}_{11}=-x_3\ell^{\alpha-1}\underset{X_1}{D}^\alpha (\frac{\mathrm{d}\bar{u}_3(x_1)}{\mathrm{d}x_1})=z\ell^{\alpha-1}\breve{\kappa}_2=-z\ell^{\alpha-1}w,_{1\breve{1}},
\label{eq:17}
\end{equation}
so 
\begin{equation}
{\sigma}_{11}=-z\ell^{\alpha-1}Ew,_{1\breve{1}},
\end{equation}
where $z=x_3$, $E$ is \textsc{Young}'s modulus, $w=\bar{u}_3(x_1)$ and $\breve{\kappa}_2(x_1)=-w,_{1\breve{1}}$. It is clear that the transverse shear stress due to flexure cannot be obtained in this way due to the third \textsc{Euler-Bernoulli} assumption - it is determined from equilibrium considerations as in the classical set-up (we further denote it by $\sigma^*_{13}$).

\begin{figure}[H]
\centering
\includegraphics[width=14cm]{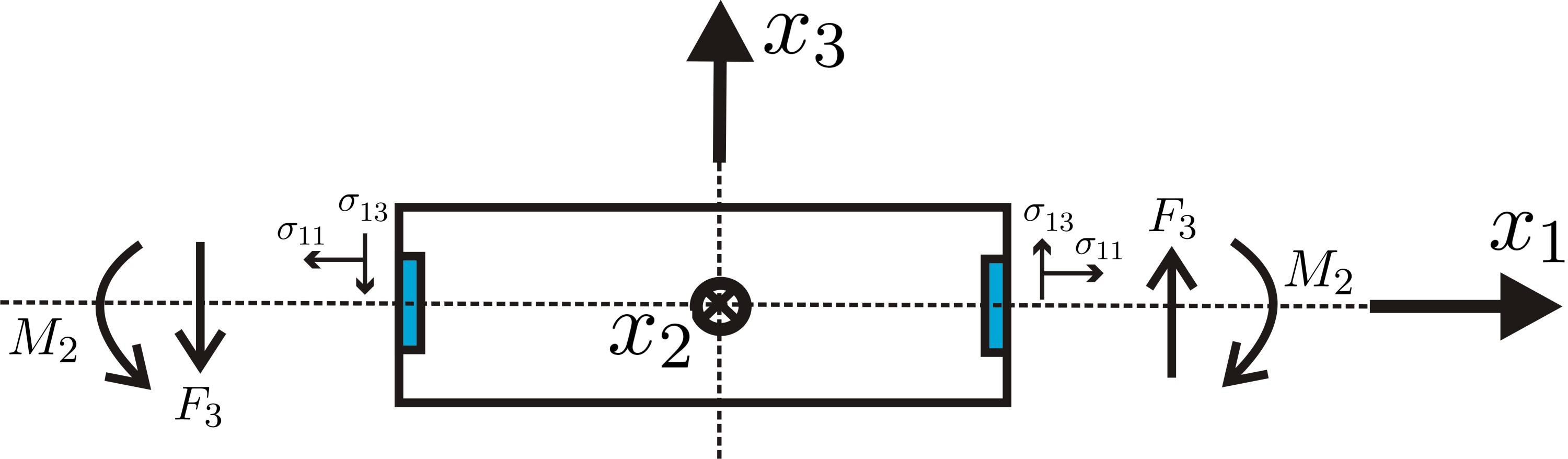}
\caption{The sectional stress resultants.}
\label{fig:sectional}
\end{figure}

Next, following Fig.~\ref{fig:sectional} the kinetic variables,
called the sectional stress resultants, are
\begin{equation}\label{eq:M2first}
M_2=\int_A \sigma_{11} x_3 \mathrm{d}A= - w,_{1\breve{1}} \int_A z^2\ell^{\alpha-1}E\mathrm{d}A,
\end{equation}
and
\begin{equation}
F_3=\int_A \sigma^*_{13} \mathrm{d}A.	
\end{equation}

Up to this point we have {three} unknowns: displacement $\bar{u}_3$, (fractional) strain $\breve{\kappa}_2$ of the beam, and stress resultant $M_1$. Thus far, we have obtained two equations which combine them, namely:
\begin{itemize}
 \item strain-displacement relation $\breve{\kappa}_2(x_1)=-\underset{X_1}{D}^\alpha (\frac{\mathrm{d}\bar{u}_3(x_1)}{\mathrm{d}x_1})$, and 
 \item constitutive relation $M_2=\breve{\kappa}_2\int_A z^2\ell^{\alpha-1}E\mathrm{d}A$.
\end{itemize}
The missing last one is derived using equilibrium considerations, so called \textsc{Newton}ian method.

\begin{figure}[H]
\centering
\includegraphics[width=14cm]{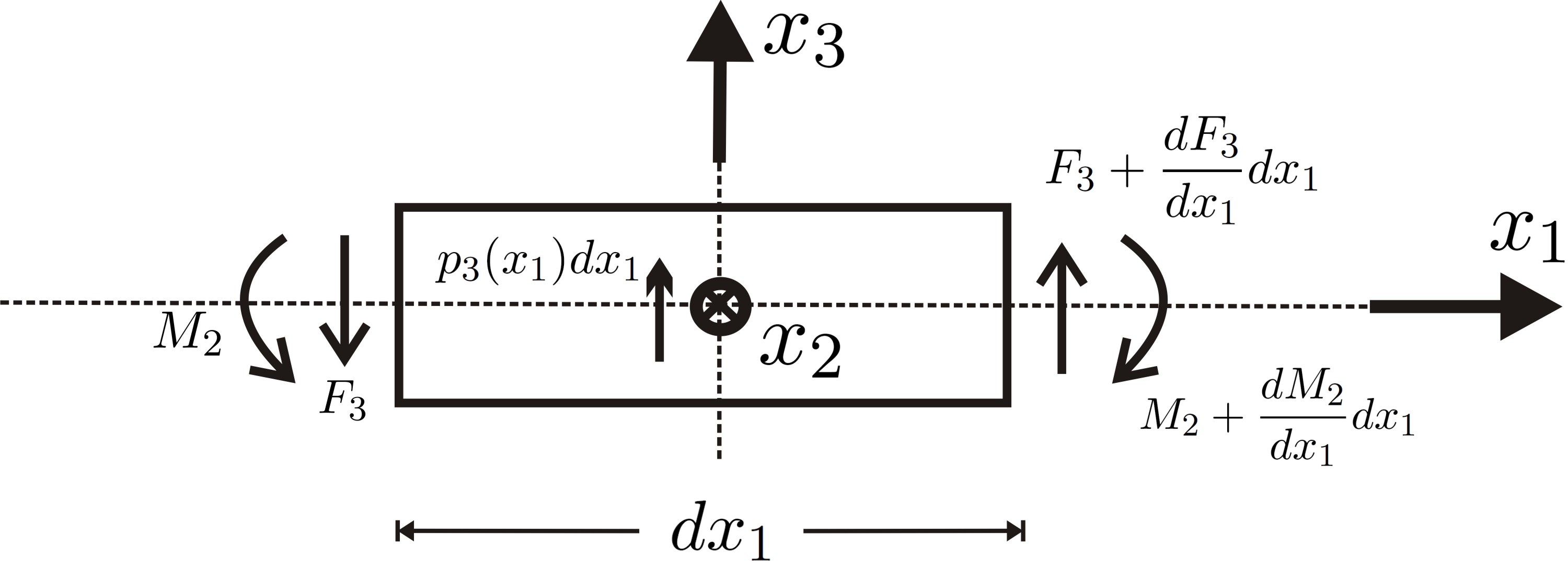}
\caption{Free body diagram for the transverse shear forces and bending moments.}
\label{fig:free}
\end{figure}

Thus, in the absence of distributed moments through the beam axis, following the free body diagram shown in Fig.~(\ref{fig:free}), we have the following relations
\begin{equation} \label{eq:F3}
\frac{\mathrm{d}F_3}{\mathrm{d}x_1}=-p_3(x_1),
\end{equation} 
and
\begin{equation} \label{eq:M2}
\frac{\mathrm{d}M_2}{\mathrm{d}x_1} - F_3 = 0,
\end{equation} 
where $p_3(x_1)$ denotes the distributed load. Next, by taking a derivative of Eq.~(\ref{eq:M2}) then introducing Eq.~(\ref{eq:F3}), the bending moment equilibrium equation is
\begin{equation} \label{eq:M2BM}
\frac{\mathrm{d}^2M_2}{\mathrm{d}x^2_1}=-p_3(x_1),
\end{equation}
which defines the third lacking equation, for the full description of beam bending {in the plane (1,3)} (in the absence of distributed moments through the beam axis).

Finally, combining Eq.~(\ref{eq:M2first}) and Eq.~(\ref{eq:M2BM}) we obtain a single equation in terms of $w$
\begin{equation}
\ell^{\alpha-1}EIw,_{1\breve{1}11}=p_3(x_1),
\end{equation}
where $I=\int_A z^2 \mathrm{d}A$. When order of fractional continua is taken $\alpha=1$ classical solution is recovered

\begin{equation}
\ell^{0}EIw,_{1111}=EI\frac{\mathrm{d}^4w}{\mathrm{d}x^4_1}=p_3(x_1).
\end{equation}

\section{Numerical study}
\begin{figure}[H]
\centering
\includegraphics[width=14cm]{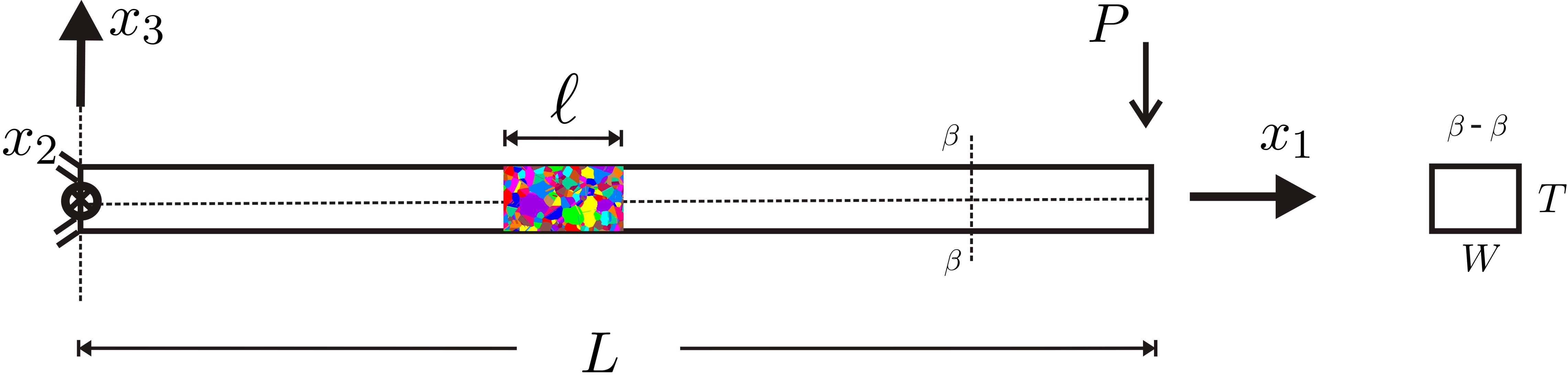}
\caption{Computational model of a micro-beam bending test.}
\label{fig:beam}
\end{figure}

{
\subsection{Bending test}\label{subsec:3.1}
The real physical conditions in micro-beam bending tests, discussed in Sect.~\ref{sec:4}, correspond to a cantilever beam model, loaded at the free end by a point load $P$ -- cf. Fig.~\ref{fig:beam}.}

For such a configuration resultant moment is known, namely
\begin{equation}
M_2(x_1)=PL-Px_1,
\end{equation}
thus Eq.~(\ref{eq:M2BM}) is satisfied. Therefore the problem is governed by (cf. Eq.~(\ref{eq:M2first}))
\begin{equation}\label{eq:MProb}
\begin{cases}
-\underset{X_1}{D}^\alpha (\frac{\mathrm{d}w}{\mathrm{d}x_1})\ell^{\alpha-1}EI=PL-Px_1,\\
w(x_1=0)=0,\\
\frac{\mathrm{d}w}{\mathrm{d}x_1}(x_1=0)=0,
\end{cases}
\end{equation} 
where $I=\frac{WT^3}{12}$.

\subsection{Computational algorithm}\label{subsec:3.2}

\begin{figure}[H]
\centering
\includegraphics[width=15cm]{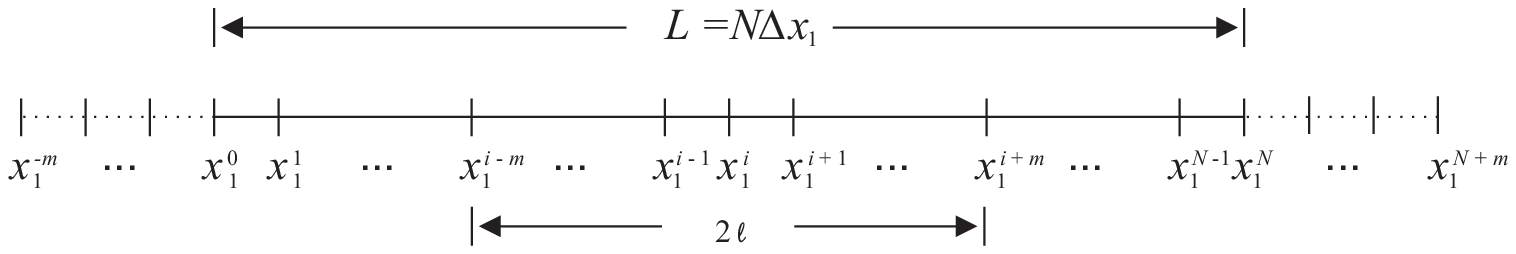}
\caption{Spatial discretization for micro-beam.}
\label{fig:grid}
\end{figure}

In this section we present a numerical scheme for the problem (\ref{eq:MProb}). We introduce the homogeneous grid of nodes (see Fig.~\ref{fig:grid}). We see that additional fictitious nodes $x_1^{-m},\ldots,x_1^{-1}$ and $x_1^{N+1},\ldots,x_1^{N+m}$ placed outside the domain $[x_1^0,x_1^N]$ are introduced. We denote a value of the beam deflection at the node $x_1^i$ as $w\left( {{x_1^i}} \right) = {w_i}$. Next, by analogy as in  {\textsc{Ciesielski \& Leszczynski} (2006)} \cite{Ciesielski2006}, {\textsc{Sumelka \& Blaszczyk} (2014)} \cite{Sumelka2014-AoM}, we assume that for all fictitious nodes on the left the beam deflections are $w_{-m}=w_{-m+1}=\ldots=w_{-1}=w_0$ and on the right the beam deflections are $w_{N+m}=w_{N+m-1}=\ldots=w_{N+1}=w_N$.
The problem of numerical solutions of equations containing simultaneously the left and right fractional derivatives has recently been drawing the attention of many authors {\textsc{Asad} et al. (2014)} \cite{Baleanu2014}, {\textsc{Blaszczyk \& Ciesielski} (2014)} \cite{Blaszczyk2014},  {\textsc{Blaszczyk} (2015)} \cite{Blaszczyk2015-rrp}, {\textsc{Cresson} et al. (2013)} \cite{Bourdin2014}, {\textsc{Xu \& Agrawal} (2014)} \cite{Xu2014}. In this paper we use a scheme which is based on the fractional trapezoidal rule \textsc{Odibat} (2006) \cite{Odibat2006}, \textsc{Blaszczyk} {et al.} (2013) \cite{Blaszczyk2013} and \textsc{Sumelka \& Blaszczyk} (2014) \cite{ Sumelka2014-AoM}. The left \textsc{Caputo} derivatives at nodes $x_1^i$ is approximated by the formula
\begin{eqnarray} \label{disc_LC1a}
 {\left. {{}_{{x_1^{i - m}}}^\mathrm{C}D_{x_1}^\alpha (\frac{\mathrm{d}w}{\mathrm{d}x_1})} \right|_{x_1 = {x_1^i}}} & = & \frac{1}{{\Gamma \left( {1 - \alpha } \right)}}\int\limits_{{x_1^{i - m}}}^{{x_1^i}} {\frac{{{w''}\left( \tau  \right)\mathrm{d}\tau }}{{{{\left( {{x_1^i} - \tau } \right)}^{\alpha }}}}} \nonumber   \\ 
& \cong & \frac{{{{\left( {\Delta x_1} \right)}^{1 - \alpha }}}}{{\Gamma \left( {3 - \alpha } \right)}}\left\{ {\left[ {{{\left( {m - 1} \right)}^{2 - \alpha}} - \left( {m + \alpha  - 2} \right){m^{1 - \alpha }}} \right]{w''}\left( {{x_1^{i - m}}} \right) + {w''}\left( {{x_1^i}} \right)} \right. \nonumber \\ 
 & + & \sum\limits_{j = i - m + 1}^{i - 1} \left. { {\left[ {{{\left( {i - j - 1} \right)}^{2 - \alpha}} - 2{{\left( {i - j} \right)}^{2 - \alpha}} + {{\left( {i - j + 1} \right)}^{2 - \alpha }}} \right]} {w''}\left( {{x_1^j}} \right)} \right\}, \nonumber \\ \end{eqnarray}
 
{where $(\cdot)''$ denotes a second derivative with respect to $x_1$}. Using the central difference formula for the second order derivatives occurring in (\ref{disc_LC1a}) we finally obtain the approximation of the left \textsc{Caputo} derivative
\begin{equation} \label{disc_LC1b}
 {\left. {{}_{{x_1^{i - m}}}^\mathrm{C}D_{x_1}^\alpha w} \right|_{x_1 = {x_1^i}}} \cong \frac{{{{\left( {\Delta x_1} \right)}^{ -1 - \alpha }}}}{{\Gamma \left( {3 - \alpha } \right)}}\sum\limits_{j = i - m - 1}^{i + 1} {{w_j}v_1\left( {i,j} \right)},
\end{equation}
where
\begin{equation}
{v_1}\left( {i,j} \right) = \left\{ {\begin{array}{*{20}{c}}
   { {{\left( {m - 1} \right)}^{2 - \alpha }} - \left( {m + \alpha  - 2} \right){m^{1 - \alpha }}} \hfill & {{\rm{for}}\;j = i - m - 1} \hfill  \\
   { {{\left( {m - 2} \right)}^{2 - \alpha }} - 4{{\left( {m - 1} \right)}^{2 - \alpha }}} \hfill & {} \hfill \\
 {  + 2\left( {m + \alpha  - 2} \right){m^{1 - \alpha }} + {m^{2 - \alpha }}} \hfill & {{\rm{for}}\;j = i - m} \hfill  \\
   { - 2{{\left( {m - 2} \right)}^{2 - \alpha }} + 5{{\left( {m - 1} \right)}^{2 - \alpha }}} \hfill & {} \hfill  \\
   { - \left( {m + \alpha  - 2} \right){m^{1 - \alpha }} - 2{m^{2 - \alpha }} + 1} \hfill & {{\rm{for}}\;j = i - m + 1 \wedge m = 2} \hfill  \\
      { {{\left( {m - 3} \right)}^{2 - \alpha }} - 4{{\left( {m - 2} \right)}^{2 - \alpha }} + 6{{\left( {m - 1} \right)}^{2 - \alpha }}} \hfill & {} \hfill  \\
   { - \left( {m + \alpha  - 2} \right){m^{1 - \alpha }} - {m^{2 - \alpha }}} \hfill & {{\rm{for}}\;j = i - m + 1 \wedge m > 2} \hfill  \\
   { {{\left( {i - j - 2} \right)}^{2 - \alpha }} - 4{{\left( {i - j - 1} \right)}^{2 - \alpha }} + 6{{\left( {i - j} \right)}^{2 - \alpha }}} \hfill & {} \hfill  \\
   { - 4{{\left( {i - j + 1} \right)}^{2 - \alpha }} + {{\left( {i - j + 2} \right)}^{2 - \alpha }}} \hfill & {{\rm{for}}\;j = i - m + 2, \ldots ,i - 2 \wedge m > 3} \hfill  \\
   {{6 - 4\cdot2^{2 - \alpha }} - 3^{2 - \alpha }} \hfill & {{\rm{for}}\;j = i - 1 \wedge m > 2 } \hfill  \\
   {{2^{2 - \alpha }} - 4} \hfill & {{\rm{for}}\;j = i} \hfill  \\
   1 \hfill & {{\rm{for}}\;j = i + 1} \hfill  \\
   0  \hfill & {{\rm{otherwise}}} \hfill  \\
\end{array}} \right..
\end{equation}
We determine the discrete form of the right fractional \textsc{Caputo} derivative in a similar way
\begin{eqnarray} \label{disc_PC1a}
 {\left. {{}_{{x_1}}^\mathrm{C}D_{x_1^{i + m}}^\alpha (\frac{\mathrm{d}w}{\mathrm{d}x_1})} \right|_{x_1 = {x_1^i}}} & = & \frac{- 1}{{\Gamma \left( {1 - \alpha } \right)}}\int\limits_{{x_1^{i}}}^{{x_1^{i+m}}} {\frac{{{w''}\left( \tau  \right)\mathrm{d}\tau }}{{{{\left( {\tau - {x_1^i}  } \right)}^{\alpha }}}}} \nonumber   \\ 
& \cong & \frac{{{{-\left( {\Delta x_1} \right)}^{1 - \alpha }}}}{{\Gamma \left( {3 - \alpha } \right)}}\left\{ {\left[ {{{\left( {m - 1} \right)}^{2 - \alpha}} - \left( {m + \alpha  - 2} \right){m^{1 - \alpha }}} \right]{w''}\left( {{x_1^{i + m}}} \right) + {w''}\left( {{x_1^i}} \right)} \right. \nonumber \\ 
 & + & \sum\limits_{j = i -  1}^{i + m + 1} \left. { {\left[ {{{\left( {j - i - 1} \right)}^{2 - \alpha}} - 2{{\left( {j - i} \right)}^{2 - \alpha}} + {{\left( {j - i + 1} \right)}^{2 - \alpha }}} \right]} {w''}\left( {{x_1^j}} \right)} \right\} \nonumber \\
 &\cong& \frac{{{{-\left( {\Delta x_1} \right)}^{ -1 - \alpha }}}}{{\Gamma \left( {3 - \alpha } \right)}}\sum\limits_{j = i  - 1}^{i + m + 1} {{w_j}v_2\left( {i,j} \right)},
 \end{eqnarray}
 where
\begin{equation}
{v_2}\left( {i,j} \right) = \left\{ {\begin{array}{*{20}{c}}
   { {{\left( {m - 1} \right)}^{2 - \alpha }} - \left( {m + \alpha  - 2} \right){m^{1 - \alpha }}} \hfill & {{\rm{for}}\;j = i + m + 1} \hfill  \\
   { {{\left( {m - 2} \right)}^{2 - \alpha }} - 4{{\left( {m - 1} \right)}^{2 - \alpha }}} \hfill & {} \hfill \\
 {  + 2\left( {m + \alpha  - 2} \right){m^{1 - \alpha }} + {m^{2 - \alpha }}} \hfill & {{\rm{for}}\;j = i + m} \hfill  \\
   { - 2{{\left( {m - 2} \right)}^{2 - \alpha }} + 5{{\left( {m - 1} \right)}^{2 - \alpha }}} \hfill & {} \hfill  \\
   { - \left( {m + \alpha  - 2} \right){m^{1 - \alpha }} - 2{m^{2 - \alpha }} + 1} \hfill & {{\rm{for}}\;j = i + m - 1 \wedge m = 2} \hfill  \\
      { {{\left( {m - 3} \right)}^{2 - \alpha }} - 4{{\left( {m - 2} \right)}^{2 - \alpha }} + 6{{\left( {m - 1} \right)}^{2 - \alpha }}} \hfill & {} \hfill  \\
   { - \left( {m + \alpha  - 2} \right){m^{1 - \alpha }} - {m^{2 - \alpha }}} \hfill & {{\rm{for}}\;j = i + m - 1 \wedge m > 2} \hfill  \\
   { {{\left( {j - i - 2} \right)}^{2 - \alpha }} - 4{{\left( {j - i - 1} \right)}^{2 - \alpha }} + 6{{\left( {j - i} \right)}^{2 - \alpha }}} \hfill & {} \hfill  \\
   { - 4{{\left( {j - i + 1} \right)}^{2 - \alpha }} + {{\left( {j - i + 2} \right)}^{2 - \alpha }}} \hfill & {{\rm{for}}\;j = i + 2, \ldots ,i + m - 2 \wedge m > 3} \hfill  \\
   {{6 - 4\cdot2^{2 - \alpha }} - 3^{2 - \alpha }} \hfill & {{\rm{for}}\;j = i + 1 \wedge m > 2 } \hfill  \\
   {{2^{2 - \alpha }} - 4} \hfill & {{\rm{for}}\;j = i} \hfill  \\
   1 \hfill & {{\rm{for}}\;j = i - 1} \hfill  \\
   0  \hfill & {{\rm{otherwise}}} \hfill  \\
\end{array}} \right..
\end{equation}
 Finally, we present the discrete form of the considered problem (\ref{eq:MProb}).
For calculation of values $w_0, w_1, \ldots, w_N$ we need to solve the system of $N+1$ linear equations.
For every grid node $x_1^i$, where $i = 0,...,N$, we can write the following equations
\begin{equation} \label{disc_FracProb}
\left\{ \begin{array}{l}
 {w_0} = 0, \\ 
 {w_1} = 0, \\ 
 \displaystyle\sum\limits_{j = i - m - 1}^{i + 1} {{w_j}{v_1}\left( {i,j} \right)}  + \displaystyle\sum\limits_{j = i - 1}^{i + m + 1} {{w_j}{v_2}\left( {i,j} \right)}  = \dfrac{{P\left( {L - i\Delta {x_1}} \right)}}{\beta }, \\ 
 \end{array} \right.
 \end{equation}
where
\begin{equation}
\beta  =  - \dfrac{{{\ell ^{\alpha  - 1}}EI}}{{{\left(\Delta {x_1}\right)^{1 + \alpha }}\Gamma \left( {3 - \alpha } \right)}}.
\end{equation} 
{The parameter $m$, in above formulae, represents the number of spatial nodes for which the fractional derivatives, Eq. (\ref{disc_LC1a}) and (\ref{disc_PC1a}), are calculated. The relation between the parameter $m$, length scale $\ell$ and step size $\Delta x_1$ is expressed by: $\Delta x_1 = \frac{\ell}{m}$. Thus, for larger values of $\ell$, $m$ should also be increased appropriately, because the convergence of solution, represented in Eq.~(\ref{disc_FracProb}), is in relation to $\Delta x_1$.}

\subsection{Benchmark results}\label{subsec:3.3}
{As in the CEBB theory, also in its fractional non-local version, the information about the cross section is introduced through the moment of inertia $I$ only. In this sense, the FEBB theory makes it possible to analyse the non-local effects through beam length $L$. In the following series of benchmark results we consider the influence of length scale $\ell$ and order of fractional continua $\alpha$ on cantilever beam bending test results.

{On the basis of the numerical scheme presented in the previous section we implemented a computer program in Maple and carried out computational simulations for different values of parameters $\alpha, \ell$ and $m$. We applied the LU decomposition method 
    in order to numerically solve the system of equations~(\ref{disc_FracProb}). In all presented {benchmark} examples we assumed $P = 1\ \mathrm{N}$, $E = 1\ \mathrm{Pa}$, $L = 1\ \mathrm{m}$, $W=T=\frac{1}{10}\ L$ and $\Delta x_1 = 0.001\ \mathrm{m}$. We denote by $w\left( L, \alpha \right)$ a value of a beam deflection at the loaded end $w\left( x_1^N \right)$ for fixed $\alpha \in \left\{{ 0.4, 0.5, 0.6, 0.7, 0.8, 0.9, 0.99} \right\}$ and by $w\left( L, 1 \right)$ a value of a beam deflection at the loaded end $w\left( x_1^N \right)$ for classical local i.e. $\alpha = 1$.} In Fig.~\ref{fig:Fig01} the numerical results are presented. {Two cases are considered: (i) the interval of fractional differentiation $2\ell$ is smaller (or equal) than the smallest beam dimension $T$, and (ii) $2\ell$ is smaller (or equal) then beam length $L$.}
    
\begin{figure}[H]
\centering
\includegraphics[width=7.6cm]{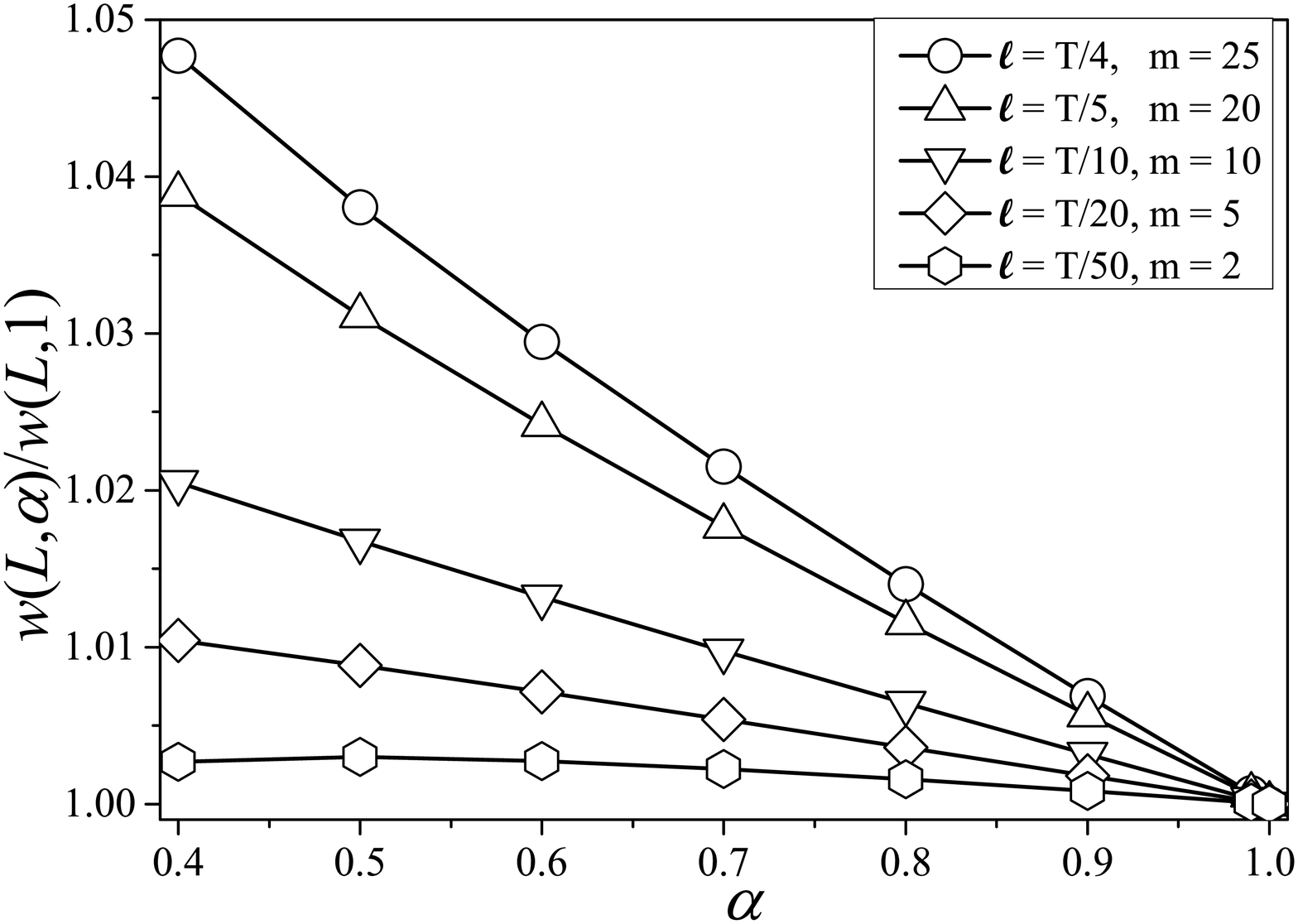}
\includegraphics[width=7.6cm]{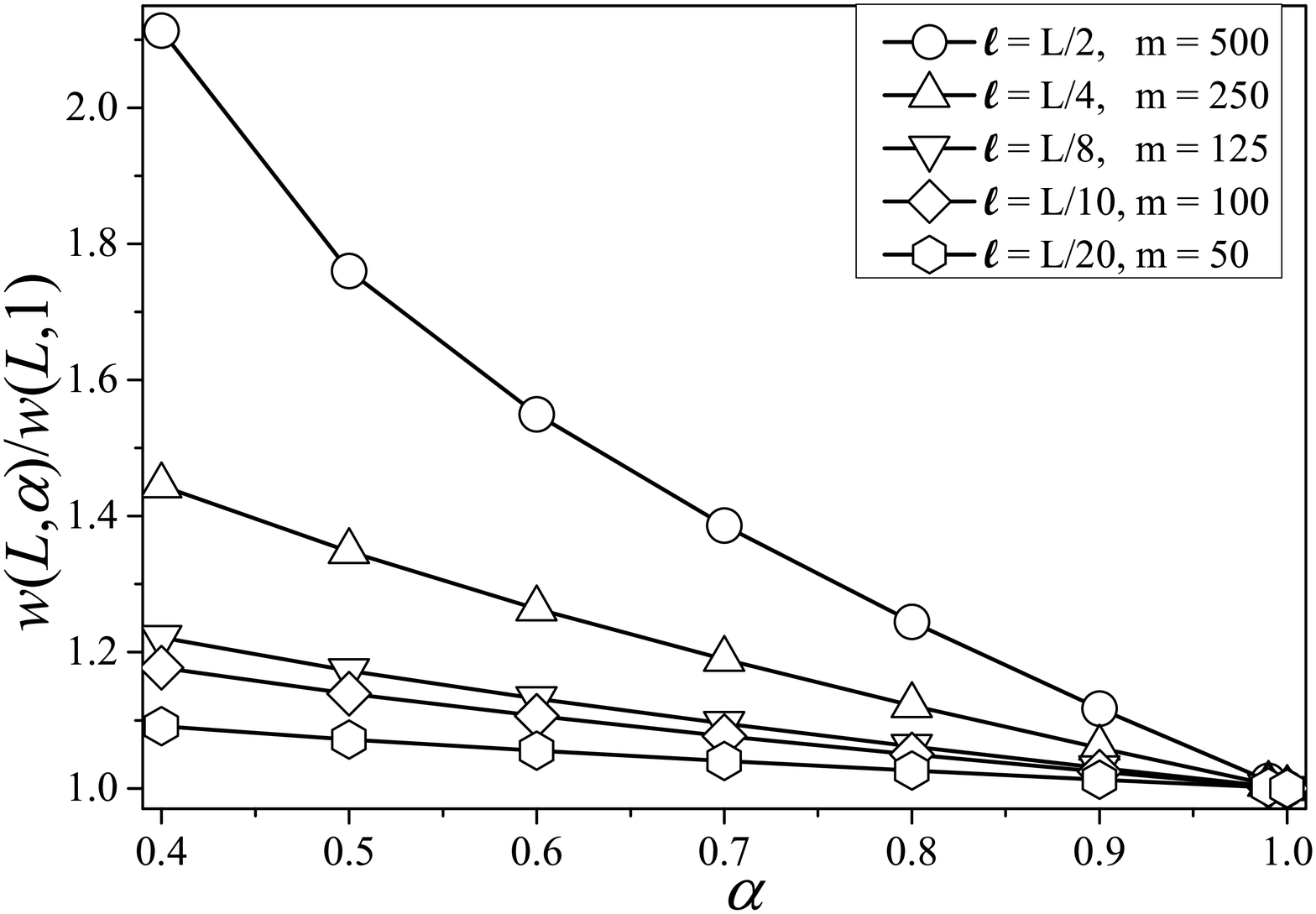}
\caption{The relation between the normalised cantilever beam deflection at loaded end vs. order of fractional continua $\alpha$ for different length scales $\ell$.Left: case (i). Right: case (ii).}
\label{fig:Fig01}
\end{figure}

We observe that when length scale $\ell$ increases in comparison to beam length $L$, the non-local effects are more vivid. {If $\ell$ approaches zero and/or $\alpha$ one, the FEBB reduces to the CEBB model}. It is also clearly seen that for smaller values of $\alpha$ the difference between a classical and fractional result increases. We can conclude that both $\ell$ and $\alpha$ control the {stiffness} of the fractional beam -- this result is crucial concerning model identification in the next Sect.~\ref{sec:4}.

\section{Experimental validation}\label{sec:4}

\subsection{Experimental set-up}

Deflection and force data from real micro-beam bending tests were recorded with an off-axis laser-reflective Mul\-ti\-view\--1000 AFM-stage from Nanonics Imaging Ltd. The system is composed of a flat scanner, consisting of a fine thread driven by piezo-elements with high-voltage power supply and a detection device working with four Photo-Sensitive Diods (PSD) interconnected as a \textsc{Wheatstone} bridge to monitor deflections of a laser-beam path. The laser is reflected in an obtuse angle from a fixed AFM-cantilever  such that the system directly monitors its deflections $w_{\text{c}}$, when deformed by the piezo uplift (referred to as separation $z$). With the knowledge of the well calibrated spring constant of the AFM-cantilever of $k_{\text{c}}=31.4\ \text{N}\ {\text{m}^{-1}}$, we were able to convert the PSD-signal  into force $P=k_{\text{c}}\ w_{\text{c}}$ (cf., Fig. \ref{fig:AFM-calibratio}). The calibration process used here is described in more detail in \textsc{Varenberg} et al. (2005) \cite{Varenberg2005} and was performed using a precise silicon normal that was provided by the PTB (Physikalisch Technische Bundesanstalt -- Braunschweig). The AFM-cantilever is a custom-built cylindrical and curved cantilever made of glass, having a tip radius of about 20~nm. In micro-beam bending tests it can be assumed that the pure AFM-data consist of a combined signal of the deflection of the AFM-cantilever and the micro-beam's deflection $w$ in the following manner: $z=w+w_{\text{c}}$. Hence, we were able to separate the deflection data of a micro-beam from the pure AFM-data in combination with the corresponding load, cf. Fig. \ref{fig:AFM-calibratio}. 

\begin{figure}[H]
\centering
\includegraphics[width=16cm]{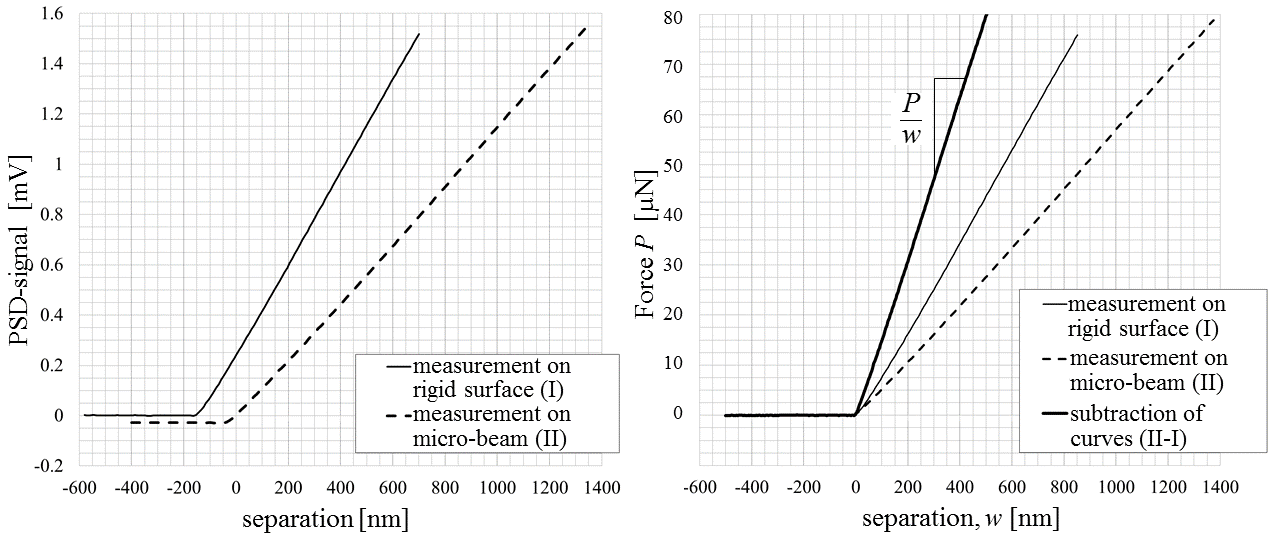}
\caption{Some exemplary original PSD-data (left) as well as shifted and converted data (right).}
\label{fig:AFM-calibratio}
\end{figure}

The ratio $P/w$ is called bending stiffness  with the classical relationship  to the elastic modulus \textit{E}, width \textit{W}, length \textit{L} and thickness \textit{T}

\begin{equation}\label{eq:AFM}
E=\frac{4L^3}{WT^3}\frac{P}{w},
\end{equation}

when the CEBB theory and a rectangular cross-section of the beam are assumed. The length and width of the samples have been measured in an optical microscope with a magnification of 500~times, whereas the thickness was taken to be the mean value from two different optical determination systems and values of a Scanning Electron Microscope (SEM). Thereby, the lengths were specified between the fixation of the cantilever on the solid glass support and the force application point of the AFM-tip, which could be varied.

\begin{figure}[h]
\centering
\includegraphics[width=6cm]{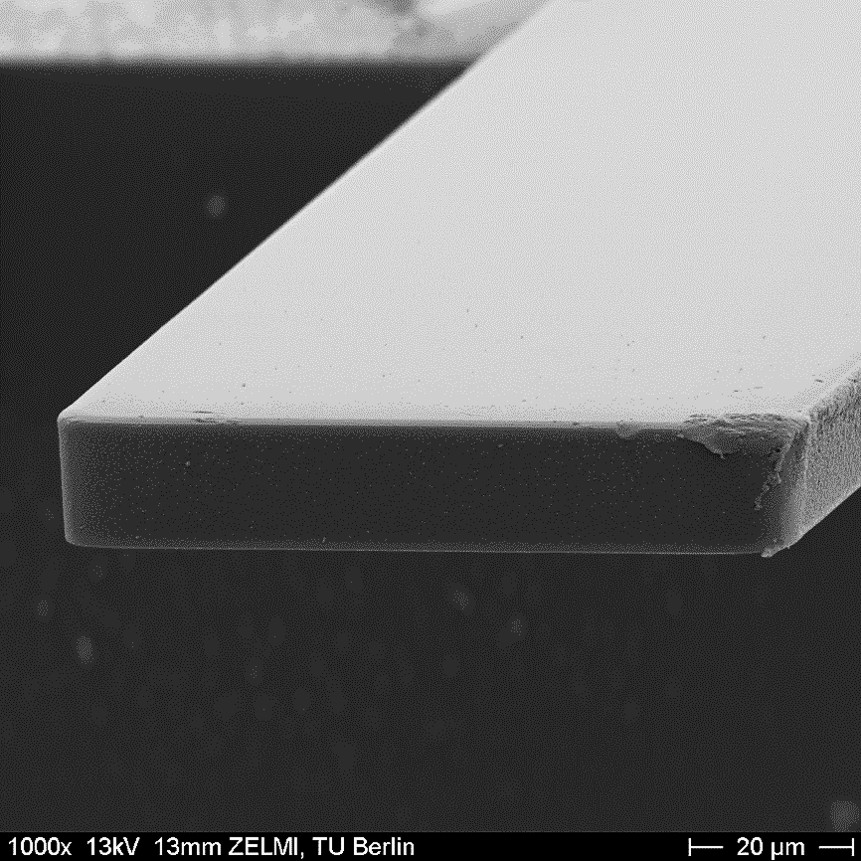}
\caption{An exemplary SEM picture of the free end of a tested micro-beam made of SU-8.}
\label{fig:SU-8_SEM}
\end{figure}
SU-8 is known as the photo-resist Nano$^{\text{TM}}$-SU-8 of the company MicroChem, used in the micro-system technology as described in \textsc{Lorenz} et al. (1997) \cite{Lorenz1997}. Manufacturing of the samples was carried out in the following steps:

\begin{itemize}[itemsep=-1pt]
\item{A 4-inch silicon wafer was coated with a thin metal film as a barrier layer.}
\item{First, the viscous SU-8 resist was dissolved in solvents. A conventional spin-coating machine was applied to disperse the resist on the metal-coated silicon wafer. The rotational speed of the coating machine determined the thickness of the homogeneous layer (between 8--40 microns).}
\item{The solvent evaporated in the rotary process in large part and the remaining part was evaporated in a subsequent drying process at temperatures of 60\textcelsius\ --\ 95\textcelsius, whereby the material finally received its rigidity.}
\item{Structuring was achieved by a Laser Direct Imager (LDI). After exposure to light an additional heat treatment was carried out at 60\textcelsius\ --\ 95\textcelsius\  to assist the chemical reaction of illumination.}
\item{By using a proper developer, the regions of exposed SU-8 were dissolved from unexposed regions.}
\item{Followed by a chemical etch process, which does not attack the SU-8, the thin metal film on the silicon wafer was dissolved and the micro-beam structures were finally peeled off.}
\item{In a last step the micro-beam structures were glued on a support made of glass and fixed on an AFM sample holder.}
\end{itemize}

Samples of different lengths and widths have been produced in different thicknesses ($T=$~8.4~$\upmu$m and 14.4~$\upmu$m) at the Fraunhofer Institute for Reliability and Microintegration, Berlin, and were tested with the AFM-technique according to Eq. \ref{eq:AFM}. 

\begin{figure}[H]
\centering
\includegraphics[width=14cm]{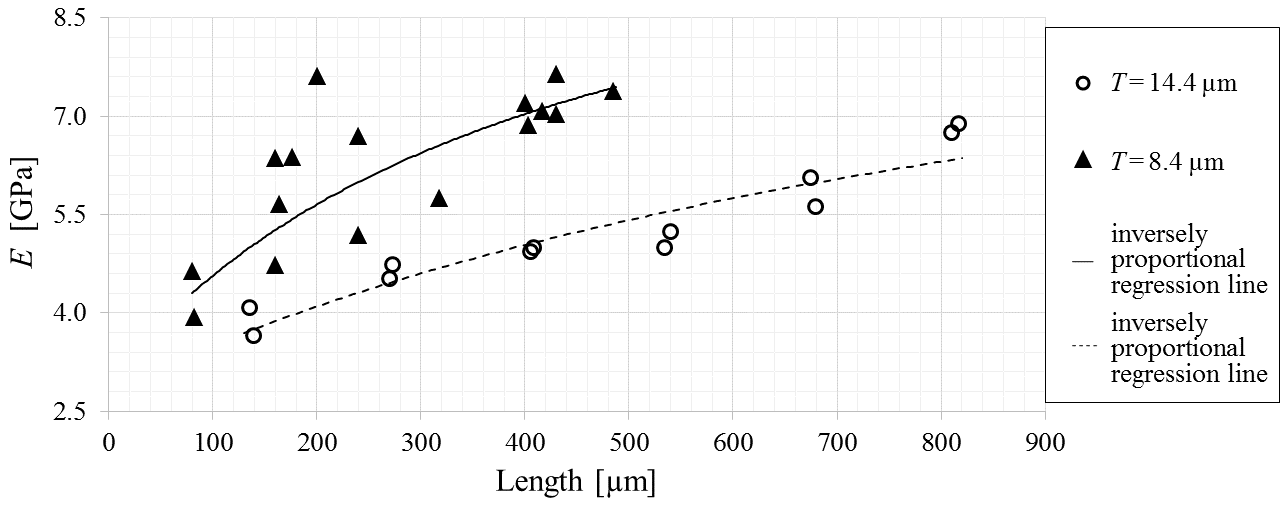}
\caption{The effect of the length of a beam made of SU-8 on the elastic modulus recorded with the AFM-technique.}
\label{fig:Length_effect}
\end{figure}
Results show that the values of elastic moduli $E$ increase when the thickness of the cantilevers is decreased. The increase of elastic modulus correlates positively with the length $L$ (see Fig. \ref{fig:Length_effect}) and the values have nearly doubled in the case of the thinnest cantilevers. The size effect, that has been detected here, manifests itself in the slopes of the partial regression lines of the experimental data which show an overall increase with decreasing beam thicknesses. On the basis of general predictions of the material surface theory (\textsc{Gurtin \& Murdoch} 1975 \cite{Gurtin1975}), regression lines were constructed as inversely proportional functions and independent of the actual thickness. We observe a dependency of $E$ on the length $L$ as well as on the thickness $T$.

\subsection{Model identification}
The discussed experimental results show a very complex behaviour of the real micro-beam. The scale effect is vivid when dimensions of rectangular prism specimen ($L,W,T$) are changed, and is more pronounced when these dimensions are comparable with an assumed real intrinsic length scale. In this sense, from the point of view of  modelling, {a 2D or even 3D description should be applied. Additionally, anisotropic length scale effects could be covered in a 2D or 3D description}. Nevertheless, we claim that the FEBB model described here is able to model the present size effects {with respect to a change of beam lengths and beam thicknesses by finding appropriate combinations of the two additional introduced parameters $\alpha$ and $\ell$, together with the type of fractional derivative (in our case the \textsc{Riesz-Caputo} one)}.
However, similar to the CEBB model, the information about the cross section is only included in the second moment of inertia (cf. Eq.~\ref{eq:MProb}). Thus, from the first sight one can expect, {that only scale effects between different micro-beam's lengths ($L$) can be covered by FEBB, but not between different thicknesses or widths} ($T$ or $W$). Fortunately, such a statement is not true {at all. In our observations, a change of both, length and cross sectional area of micro-beams can be covered by FEBB formulation, by means of changing the order of fractional continuum $0\mkern-3mu<\mkern-2mu\alpha\mkern-3mu<1$}.
In Fig.~\ref{fig:Fig03} the comparison of the experimental results (cf. Fig.~\ref{fig:Length_effect}) and the results of the FEBB model are presented. {Non-local moduli $E_{\text{NL}}$ have been calculated according to Eq.~\ref{eq:AFM}, utilising numerical deflections $w\left( L, \alpha \right)$}. We observe, that the length scale in our FEBB model should be $\ell=60~\upmu$m, whereas \textsc{Young}'s modulus should be $E=6.9$~GPa. The effect of a change of cross sections is modelled using a different order of fractional continua, namely: for $T=8.4~\upmu$m and $W=82~\upmu$m we have $\alpha=0.8$; and for $T=14.4~\upmu$m and $W=122~\upmu$m we have $\alpha=0.4$. It should also be emphasised, that the experimental result, that when the length $L$ of micro-beams is growing, {the non-local elastic modulus $E_{\text{NL}}$ converges to a single value}. {This gives evidence to the previously expected behaviour of the FEBB model, that if the dimensions of beams become significantly larger than the length scale ($L,W,T\mkern+2mu \gg \ell$), the non-local modulus $E_{\text{NL}}$ converges to the classical elastic modulus $E$}.
\begin{figure}[H]
\centering
\includegraphics[width=14cm]{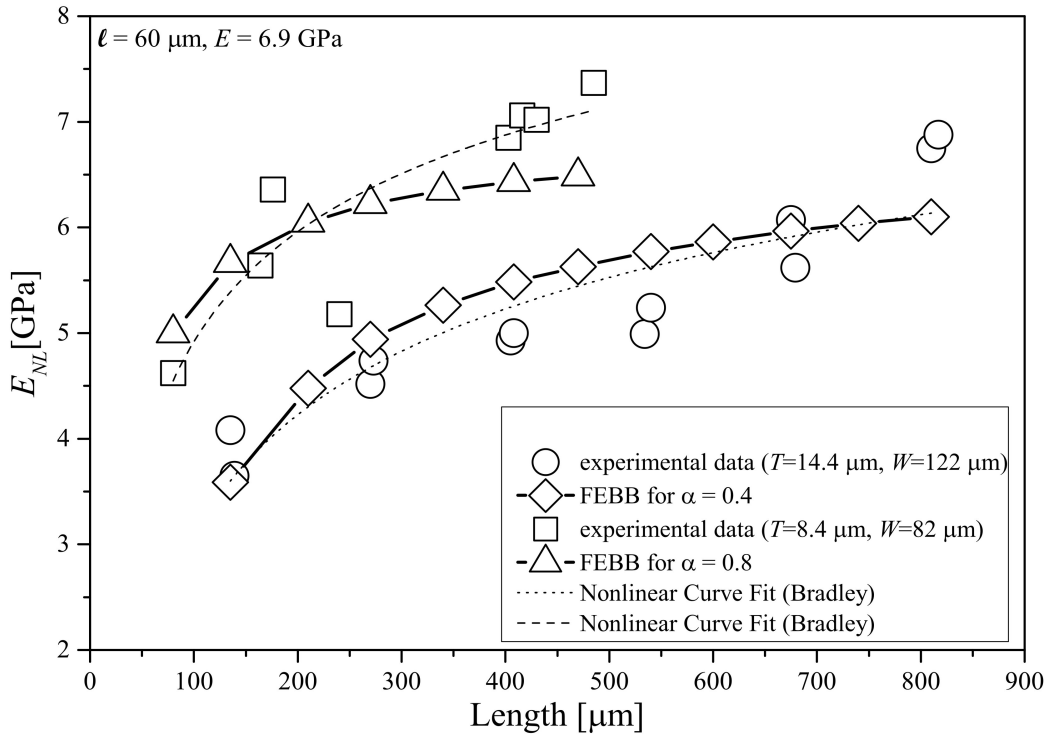}
\caption{Comparison of the experimental results with the FEBB model.}
\label{fig:Fig03}
\end{figure}

\section{Conclusions}
In the present work, the non-local \textit{fractional Euler-Bernoulli beam} theory is formulated as a generalisation of classical \textsc{Euler-Bernoulli} beams, utilising fractional calculus. The model is implemented using codes developed by the authors and identified based on AFM experiments concerning bending of micro-beams made of the polymer SU-8, where scale effect is severely manifested. The fractional \textsc{Euler-Bernoulli} beam model introduces two additional parameters $\alpha$ and $\ell$ together with the type of fractional derivative (in the discussed case the \textsc{Riesz-Caputo} one). It is shown, that this new model is able to give, in a qualitative and quantitative manner, good approximation of the revealed experimental results.

\section*{Acknowledgement}
The present work is partially supported by (Deutsche Forschungsgemeinschaft) DFG under Grant MU 1752/33-1 and the National Centre for Research and Development (NCBiR) under Grant No. UOD-DEM-1-203/001. The authors would like to thank the Fraunhofer Institute for Reliability and Microintegration Berlin for sample preparation and PTB-Braunschweig for AFM calibration.

\end{document}